\documentstyle[11pt]{amsart}

\begin{document}

%\catchline{}{}{}

\title{FORMS ON VECTOR BUNDLES OVER COMPACT REAL HYPERBOLIC MANIFOLDS}
\author{ A. A. BYTSENKO, A. E. GON\c CALVES} 
\address{Departamento de F\'isica, Universidade Estadual de Londrina,
Caixa Postal 6001, Londrina--Parana, Brazil}

\author{F. L. WILLIAMS}
\address{ Department of Mathematics and Statistics, University of 
Massachusetts, 
Amherst, Massachusetts 01003--4515, USA}

\maketitle

%\pub{Received 15 January 2003}{}

\begin{abstract}

We study gauge theories based on abelian $p-$ forms on real compact
hyperbolic manifolds. The tensor kernel trace formula and the
spectral functions associated with free generalized gauge fields
are analyzed.

%\keywords{Differential forms and gauge fields; hyperbolic geometry.}
\end{abstract}

\section{Introduction}  %) A SECTION HEADING

Skew symmetric tensor fields play an important role in quantum field 
theory, supergravity, and string theory, where they naturally couple to
two--form connections. The two--form in string theory (the so called 
$B-$ field)
is described at low energies by a Maxwell type gauge theory which can be 
extended to higher $p-$ forms. 
The abelian two--forms are closely related to 
the theory of gerbes which play a role in string 
theory \cite{Sharpe1,Sharpe2,Kalkkinen1,Kalkkinen2,Freed,Boer,Keurentjes}. 
Such forms can be understood as a connection on an abelian gerbe.

In the abelian case (which will be considered
in this paper) the self--dual two--form can be easily reduced to the abelian 
one--form gauge field. Generally, the covariant quantization of skew
symmetric tensor fields has met difficulties with ghost counting
and BRST--transformations.
In the framework of functional integration the
covariant quantization of free generalized gauge fields -- $p-$ forms
and the BRST--transformations have been obtained in
Ref. 8.
We note the topological anomaly which can be computed using the 
Atiyah--Singer index theorem. Typically the anomaly is geometric, i.e.
it is a smooth line bundle with hermitian metric and a compatible connection.
The geometric anomaly (the pfaffian line bundle with metric and connection)
may be computed using differential $K-$ theory, a version of $K-$
theory which includes differential forms as curvatures \cite{Lott,Freed1}.

In this paper we show that for $p-$ forms on real hyperbolic
manifolds the variables contribution reduces to a simple alternating
sum of forms. We present a decomposition of the Hodge Laplacian
and the tensor kernel trace formula associated to free generalized 
gauge fields -- $p-$ forms. 
The main ingredient required is a type
of differential form structure on the physical, auxiliary, or ghost
variables. We consider spectral functions on hyperbolic
manifolds associated with physical degrees of freedom of the Hodge--de
Rham operators on $p-$ forms. 

\section{Exterior Forms of Real Hyperbolic Spaces}

We shall work with an $n-$ dimensional compact real hyperbolic space 
$X$ 
with universal covering $M$ and fundamental group $\Gamma$. We can 
represent $M$ as the symmetric space $G/K$, where $G=SO_1(n,1)$ and 
$K=SO(n)$ is a maximal compact subgroup of $G$. Then we regard $\Gamma$ 
as a discrete subgroup of $G$ acting isometrically on $M$, and we take 
$X$ to be the quotient space by that action: $X=\Gamma\backslash M=
\Gamma\backslash G/K$. Let $\tau$ be an irreducible representation of 
$K$ on a complex vector space $V_\tau$, and form the induced 
homogeneous
vector bundle $G\times_K V_\tau$ (the fiber product of $G$ with 
$V_\tau$
over $K$) $\longrightarrow M$ over $M$. Restricting the $G$ action to 
$\Gamma$ we obtain the quotient bundle $E_\tau=\Gamma\backslash
(G\times_KV_\tau)\longrightarrow X=\Gamma\backslash M$ over $X$. The 
natural Riemannian structure on $M$ (therefore on $X$) induced by 
the Killing form $(\;,\;)$ of $G$ gives rise to a connection Laplacian 
${\frak L}$ on $E_\tau$. If $\Omega_K$ denotes the Casimir operator of 
$K$ -- that is
\begin{equation}\label{01}%%%%%%%
\Omega_K=-\sum y_j^2,
\end{equation}
\\
for a basis $\{y_j\}$ of the Lie algebra ${\frak k}_0$ of $K$, where 
$(y_j\;,y_\ell)=-\delta_{j\ell}$, then 
$\tau(\Omega_K)=\lambda_\tau{\mathbf 1}$ 
for a suitable scalar $\lambda_\tau$. Moreover for the Casimir operator 
$\Omega$ of $G$, with $\Omega$ operating on smooth sections 
$\Gamma^\infty E_\tau$ of $E_\tau$ one has 

\begin{equation}\label{02}%%%%%%%
{\frak L}=\Omega-\lambda_\tau{\mathbf 1}\;;
\end{equation}
\\
see Lemma 3.1 of Ref. 11. For $\lambda\geq 0$ let 

\begin{equation}\label{03}%%%%%%%
\Gamma^\infty\left(X\;,E_\tau\right)_\lambda=
\left\{s\in\Gamma^\infty E_\tau\left|-{\frak L}s=\lambda s\right.
\right\}
\end{equation}
\\
be the space of eigensections of ${\frak L}$ corresponding to $\lambda$. 
Here we note that since $X$ is compact we can order the spectrum of 
$-{\frak L}$ by taking $ 0=\lambda_0<\lambda_1<\lambda_2<\cdots$; 
$\lim_{j\rightarrow\infty}\lambda_j=\infty$. We shall focus on the more 
difficult (and more interesting) case when $n=2k$ is even, and we shall 
specialize $\tau$ to be the representation $\tau^{(p)}$ of $K=SO(2k)$ 
on $\Lambda^p {\Bbb C}^{2k}$, say $p\neq k$. The case when $n$ is 
odd will be dealt with later. It will be convenient moreover to work 
with the normalized Laplacian ${\frak L}_p=-c(n){\frak L}$ 
where $c(n)=2(n-1)=2(2k-1)$. ${\frak L}_p$ has spectrum 
$\left\{c(n)\lambda_j\;,m_j\right\}_{j=0}^\infty$ where the 
multiplicity
$m_j$ of the eigenvalue $c(n)\lambda_j$ is given by 

\begin{equation}\label{04}%%%%%%%
m_j={\rm 
dim}\;\Gamma^\infty\left(X\;,E_{\tau^{(p)}}\right)_{\lambda_j}\;.
\end{equation}

\subsection{Quantum dynamics of exterior forms}

Let $T_{j_1j_2...j_k}$ be a skew--symmetric tensor of $(0,k)-$ type,
i.e.\,\,\,\, $T_{\sigma(j_1,...,j_k)}\stackrel{def}{=}\,\,\,
{\rm sgn}(\sigma)T_{j_1,j_2,...,j_k}$, where ${\rm sgn}(\sigma)=\pm 1$ 
is the sign of a permutation $\sigma$. 
The exterior differential $p-$ form is

\begin{equation}
\omega_p = (1/p!)\sum_{j_1,...,j_p}T_{j_1j_2...j_p}
dx^{j_1}\wedge ... \wedge dx^{j_p}
\mbox{.}
\end{equation}
\\
Here $\wedge$ is the exterior product,
$dx^j$ are the basis one--forms, and
$j= 1,2, ..., n$ (${\rm dim}\,M=n$).
Let $\Lambda^*(M)\equiv \oplus_{p=0}^{n}\Lambda^p$ be the graded Cartan 
exterior algebra of differential forms; $\Lambda^p$ is the space of all
$p-$ forms on $M$. Let $(*T)$ denote a skew symmetric tensor of 
$(0,n-k)$ 
type, i.e.

\begin{equation}
(*T)_{j_{k+1...j_n}} = (1/k!)\sqrt{|{\rm g}|}
\varepsilon_{j_1...j_n}T^{j_1... j_k}\,\,,\,\,\,\,\,\,\,\,\,\,
\,\,\,\,\,
T^{j_1... j_k} ={\rm g}^{j_1\ell_1}\cdot\cdot\cdot
{\rm g}^{j_k\ell_k}T_{\ell_1... \ell_k}
\mbox{,}
\end{equation}
\\
where $\varepsilon_{j_1...j_n} = \pm 1$ for 
${\rm sgn}(j_1...j_n)= \pm 1$ is the Levi--Civita tensor density, 
and the metric ${\rm g}_{j\ell}$
(an external gravitational field) has the signature $(+,+,...,+)$.
In local coordinates the exterior differential 
$d: \Lambda^p \longrightarrow \Lambda^{p+1}$, and the co--differential
$\delta: \Lambda^p\longrightarrow \Lambda^{p-1}$ take respectively the 
form:

$$
d\omega = (1/p!)\sum_{j_1,j_2,...,j_{p+1}}
\frac{\partial T_{j_2...j_{p+1}}}{\partial x^{j_1}}
dx^{j_1}\wedge ... \wedge dx^{j_{p+1}}
\mbox{,}
$$

\begin{equation}
\delta\omega = - (1/(p-1)!)\sum_{j_1,j_2,...,j_{p-1}}
\frac{\partial T_{j_1...j_p}}{\partial x_{j_1}}
dx^{j_2}\wedge ... \wedge dx^{j_p}
\mbox{.}
\end{equation}
\\
From last equations it is easy to prove the following properties for 
operators 
and forms: $dd=\delta\delta=0$,\, $\delta = (-1)^{np+n+1}*d*$,\,
$**\omega_p = (-1)^{p(n-p)}\omega_p$. Let $\alpha_p,\, \beta_p$ be $p-$
forms; then the invariant inner product is defined by
$(\alpha_p, \beta_p)\stackrel{def}{=}\int_M \alpha_p\wedge*\beta_p$.
The operators $d$ and $\delta$ are adjoint to each other with respect 
to this inner product for $p-$ forms: 
$(\delta\alpha_p, \beta_p) = (\alpha_p, d\beta_p)$.
\\
\\
In quantum field theory the Lagrangian associated with $\omega_p$
takes the form:
$L=d\omega_p\wedge *d\omega_p$ (gauge field);\,
$L=\delta\omega_p\wedge*\delta\omega_p$ (co--gauge field).
The Euler--Lagrange equations supplied with the gauge give
${\frak L}_p\omega_p =0\,,\,\,\delta\omega_p =0$ (Lorentz gauge);\,
${\frak L}_p\omega_p =0\,,\,\, d\omega_p =0$ (co--Lorentz gauge).
These Lagrangians give possible representation of tensor fields or
generalized abelian gauge fields. The two representations of tensor
fields are not completely independent. Indeed, there is a duality
property in the exterior calculus which gives a connection between 
star--conjugated gauge tensor fields and co--gauge fields. The gauge
$p-$ forms map into the co--gauge $(n-p)-$ forms under the action of 
the Hodge operator $(*)$.
From the Hodge theory we have the orthogonal 
decomposition of $p-$ forms 
\begin{equation}
\omega_p=\delta\omega_{p+1}+d\omega_{p-1}+Har_p
\mbox{,}
%\eqno{(12.32)}
\end{equation}
\\
with $Har_p$ being a harmonic $p-$ form. 
It is known that the $L^2$ harmonic $p-$ form $Har_p^{(2)}$ appear on even 
real hyperbolic manifolds only. The following result holds:

\medskip
\par \noindent
{\bf Proposition 1:}\,\,\,\, ({\it Ref. 12, p. 373}).\,\,\,\, 
{\em The manifold ${\Bbb H}^n$ admits $L^2$ harmonic $p-$ forms
if and only if $n=2p$. For even dimensional real hyperbolic manifolds
the space of $L^2$ harmonic $p-$ forms is infinite dimensional.}
\\
\\
One can consider the $L^2-$ de Rham complex:

\begin{equation}
0\longrightarrow
\Lambda^0 (M)\stackrel{d_0}{\longrightarrow} ...
\longrightarrow \Lambda^p(M)\stackrel{d_p}{\longrightarrow} 
\Lambda^{p+1}(M)\stackrel{d_{p+1}}{\longrightarrow} ...
\longrightarrow \Lambda^n(M) \longrightarrow 0 
\mbox{,}
\end{equation}
\\
and its associated $L^2-$ cohomology

\begin{equation}
H^p(M)=\frac{{\rm ker}\,(\Lambda^p(M)\stackrel{d_p}{\longrightarrow} 
\Lambda^{p+1}(M))}
{{\rm range}\,(d_{p-1}\Lambda^{p-1}(M))}
\mbox{.}
\end{equation}
\\
The theorem of Kodaira Ref. 13, p. 165 gives the following injection:

\begin{equation}
{Har}_p^{(2)}(M)\stackrel{j}{\longrightarrow} H^p(M)
\mbox{.}
\end{equation}
\\
The map (injection) $j$ is an isomorphism if and only if $d_{p-1}$ has
closed range. If $j$ is not an isomorphism, then $j$ has infinite 
dimensional
co--kernel. The associated Laplacian ${\frak L}_p$ has closed range if 
and only
if $d_p$ and $d_{p-1}$ have closed range Ref. 14, p. 446.

\section{The Trace Formula Applied to the Tensor Kernel}

Since $\Gamma$ is torsion free, each
$\gamma\in\Gamma-\{1\}$ can be represented uniquely as some power of a 
primitive
element $\rho:\gamma=\rho^{j(\gamma)}$ where $j(\gamma)\geq 1$ is an 
integer and
$\delta$ cannot be written as $\gamma_1^j$ for $\gamma_1\in \Gamma$, 
\,\, $j>1$ an
integer. Taking $\gamma\in\Gamma$, $\gamma\neq 1$, one can find 
$t_\gamma>0$
and $m_{\gamma}\in {\frak M} \stackrel{def}{=}\{m_{\gamma}\in K | 
m_{\gamma}a=
am_{\gamma}, \forall a\in A\}$ such that $\gamma$
is $G$ conjugate to $m_\gamma\exp(t_\gamma H_0)$, namely for some
${\rm g}\in G, \,{\rm g}\gamma {\rm g}^{-1}=m_\gamma\exp(t_\gamma 
H_0)$ ; that is, $\gamma$ is $G-$ conjugate to 
$m_\gamma\exp(t_\gamma H_0)$ and
$m_\gamma\in SO(n-1)$. For $\mbox{Ad}$ denoting the
adjoint representation of $G$ on its complexified Lie algebra, one can 
compute $t_\gamma$ as follows \cite{Wallach1}:

\begin{equation}\label{09b}%%%%%%%
e^{t_\gamma}={\rm max}\left\{
|c|\left|c= {\rm an\,\,\, eigenvalue\,\,\, of\,\,\,\, Ad}
(\gamma):{\rm g}\rightarrow {\rm g}\right.
\right\}\;,
\end{equation}
\\
Also $\gamma=\delta^{j(r)}$ where $j(r)\geq 1$ is a whole number
and $\delta\in\Gamma-\{1\}$ is a primitive element; ie. $\delta$ can not 
be 
expressed as $\gamma_1^j$ for some $\gamma_1\in\Gamma$ and some whole
number $j>1$. The pair $(j(\gamma),\;\delta)$ is uniquely determined by 
$\gamma\in \Gamma-\{1\}$. These facts are known to follow since $\Gamma$
is torsion free. 

Let $a_0, n_0$
denote the Lie algebras of $A, N$ in an Iwasawa decomposition $G=KAN$. 
The complexified Lie algebra ${\rm g}={\rm g}_0^{\Bbb C}
={so}(2k+1, {\Bbb C})$ of $G$ is of Cartan type $B_k$ with the Dynkin diagram

\begin{equation}
  \underbrace{\bigcirc-\bigcirc-\bigcirc \cdots \bigcirc-\bigcirc}_{2k\,
   {\rm  nodes}} = \bigcirc\,\,.
%%{(3.1)}
\end{equation}
\\
Since the rank of $G$ is one,
$\dim a_0=1$ by definition, say $a_0={\Bbb R}H_0$ for a suitable basis 
vector $H_0$:

\begin{equation}
H_0 = \left[ \begin{array}{ll}
0\,\,0\,\,\,\,\,\,\,\,\:\: .\,\,.\,\,.\,\,\,\,\,\,\,\,\:\: 0\,\, &1\\
0                                                &0\\
.                                                &.\\
.                                                &.\\
.                                                &.\\
0                                                &0\\
1\,\,0\,\,\,\,\,\,\,\,\:\: .\,\,.\,\,.\,\,\,\,\,\,\,\,\:\: 0\,\, &0  
\end{array} \right]
\mbox{,}
\end{equation}
\\
\\
is a $(k+1)\times(k+1)$ matrix.
By this choice we have the normalization $\beta(H_0)=1$, where
$\beta: a_0\rightarrow{\Bbb R}$ is the positive root which defines 
$n_0$; for more detail see Ref. 15.
Define $C(\gamma)$ on $\Gamma-\{1\}$ by 

\begin{equation}
C(\gamma)\stackrel{def}=e^{-\rho_0t_\gamma}|\mbox{det}_{n_0}\left(\mbox{Ad}
(m_\gamma
e^{t_\gamma H_0})^{-1}-1\right)|^{-1}\mbox{.}
\end{equation}
\\ 
Lastly, let $C_\Gamma\subset\Gamma$ be a complet set of representations 
in $\Gamma$ of its conjugacy classes. This means that any two elements
in $C_\Gamma$ are non--conjugate, and any $\gamma\in \Gamma$ is 
$\Gamma-$
conjugate to some element $\gamma_1\in C_\Gamma:x\gamma 
x^{-1}=\gamma_1$
for some $x\in \Gamma$. The reader may consult the appendix of 
\cite{Williams} for further structural data concerning the Lie group 
$SO_1(n,1)$ (and other rank 1 groups). Note that the Killing 
form $(\ ,\ )$ is given by $(x,y)=(n-1){\rm trace }(xy)$ for $x,\;y\in
{\rm g}_0$.
The standard systems of positive roots $\triangle^{+},\triangle^{+}_\ell$ 
for ${\rm g}$ and ${\frak k}={\frak k}^ {\Bbb C}_0$ --- 
the complexified Lie algebra of $K$, with 
respect to a Cartan subgroup $H$ of $G$,\, $H\subset K$, are given by

\begin{equation}
   \triangle^{+}=\{\varepsilon_i|1\leq i\leq k\}\bigcup 
   \triangle^{+}_\ell\,,
\end{equation}
where

\begin{equation}
\triangle^{+}_\ell=\{\varepsilon_i\pm \varepsilon_j|1\leq i<j\leq k\},
	%%{(3.3)}
\end{equation}
and

\begin{equation}
       \triangle^{+}_k \stackrel{def}{=}\{\varepsilon_i|1\leq i\leq k\}
  	%%{(3.4)}
\end{equation}
is the set of positive non--compact roots. Here

\begin{equation}
(\varepsilon_i,\, \varepsilon_j)=\frac{\delta_{ij}}{(H_0, H_0)} =
\frac{\delta_{ij}}{2(2k-1)}\,,
\end{equation}

\begin{equation}
(\varepsilon_i \pm \varepsilon_j,\,\,\varepsilon_i \pm \varepsilon_j)
= \frac{1}{2k+1}\,\,\,,\,\,\, i<j\,;\,\,\,\,\, {\rm i.e.} \,\,\,\,\,
(\alpha, \alpha) = \frac{1}{2k-1}\,\,  \forall \alpha\in \triangle^{+}_k
\mbox{.}
\end{equation}
\\
Let $\tau = \tau^{(j)}=$ representation of $K$ on $\Lambda^j{\Bbb C}^{2k}$,
$\Lambda_{\tau^{(j)}}=\Lambda_j =$ highest weight of $\tau$ is

\begin{equation}
\left[ \begin{array}{ll}
\varepsilon_1 + \cdots + \varepsilon_j\,\,\,\,\,\, &{\rm if}\,\,\, j\leq k\\
\varepsilon_1 + \cdots + \varepsilon_{2k-j}\,\,\,\,\,\,&{\rm if}\,\,\, j>k
\end{array} \right]
\mbox{.}
\end{equation}
\\
Writing $(\Lambda_j, \Lambda_j+2\delta_k)= 
(\Lambda_j, \Lambda_j) + (\Lambda_j, 2\delta_k)$,\,\, 
$\delta_k = \sum_{i=1}^k (k-i)\varepsilon_i$, for $j\leq k$ we have

\begin{equation}
(\Lambda_j,\, \Lambda_j) = 
\left(\sum_{p=1}^j\varepsilon_p\,,\,\sum_{q=1}^j\varepsilon_q\right)
= \sum_{p, q =1}^j (\varepsilon_p,\,\varepsilon_q)
=\sum_{p, q =1}^j (\varepsilon_p,\,\varepsilon_p)
= \frac{j}{(H_0, H_0)}\,\,
\mbox{,}
\end{equation}

$$
(\Lambda_j, 2\delta_k) =
\left(\sum_{p=1}^j\varepsilon_p,\,\, 2\sum_{i=1}^j(k-i)\varepsilon_i +
2\sum_{i=j+1}^k(k-i)\varepsilon_i\right)
= 2\sum_{p=1}^j(\varepsilon_p,\,\, (k-p)\varepsilon_p)
$$

\begin{equation}
= \frac{2kj}{(H_0, H_0)} -2\sum_{p=1}^jp(\varepsilon_p, \varepsilon_p)
= \frac{2kj}{(H_0, H_0)} -\frac{2j(j+1)}{(H_0, H_0)^2}\,\,
\mbox{.} 
\end{equation}
Therefore,

\begin{equation}
(\Lambda_j, \Lambda_j+2\delta_k) =
\frac{j+2kj-j(j+1)}{(H_0, H_0)}
+ \frac{2kj-j^2}{(H_0, H_0)}\,\,
\mbox{.}
\end{equation}
In the case $j>k$ we have

\begin{equation}
(\Lambda_j, \Lambda_j)=
\left(\sum_{p=1}^{2k-j}\varepsilon_p, 
\sum_{q=1}^{2k-j}\varepsilon_q\right)
= \sum_{p=1}^{2k-j}(\varepsilon_p, \varepsilon_p) 
=\frac{2k-j}{(H_0, H_0)}\,\,
\mbox{,}
\end{equation}

$$
\!\!\!\!\!\!\!\!\!\!\!\!\!\!\!\!\!\!\!\!\!\!\!\!\!\!\!\!\!\!
\!\!\!\!\!\!\!\!\!\!\!\!\!\!\!\!\!\!\!\!\!\!\!\!\!\!\!\!\!\!
\!\!\!\!\!\!\!\!\!\!\!\!\!\!\!\!\!\!\!\!\!\!\!\!\!\!\!\!\!\!
\!\!\!\!\!\!\!\!\!\!\!\!\!\!\!\!\!\!\!\!\!\!\!\!\!\!\!\!\!\!
(\Lambda_j,\, 2\delta_k) =
2\left(\sum_{p=1}^{2k-j}\varepsilon_p, 
\sum_{i=1}^{k}(k-i)\varepsilon_i\right)
$$

$$
\!\!\!\!\!\!\!\!\!\!\!\!\!\!\!\!\!\!\!\!\!\!\!\!\!\!\!\!\!\!
\!\!\!\!\!\!\!\!
=2\left(\sum_{p=1}^{2k-j}\varepsilon_p, 
\sum_{i=1}^{2k-j}(k-i)\varepsilon_i +
\sum_{i=2k-j+1}^{k}(k-i)\varepsilon_i \right)
$$

$$
=2\left(\sum_{p=1}^{2k-j}\varepsilon_p, 
\sum_{i=1}^{2k-j}k\varepsilon_i -
\sum_{i=1}^{2k-j}i\varepsilon_i \right)
=\frac{2k(2k-j)}{(H_0, H_0)} - 
2\sum_{i=1}^{2k-k}i(\varepsilon_i, \varepsilon_i)
$$

\begin{equation}
\!\!\!\!\!\!\!
= \frac{2k(2k-j) - (2k-j)(2k-j+1)}{(H_0, H_0)}
= \frac{(2k-j)(j-1)}{(H_0, H_0)}\,\,
\mbox{.}
\end{equation}
\\
\vspace{0.2cm}
Thus for $\Lambda_j = \triangle_{\ell}^{+}-$ highest weight of

$K=SO(2k)$ on $\Lambda^j{\Bbb C}^{2k}$, we have 
$(\Lambda_j, \Lambda_j + 2\delta_k) = (2kj-j^2)/(H_o, H_0) = 
(2kj-j^2)/(2(2k-1))$ for $0\leq j\leq 2k$.
\\
\\
Note that since we have specialized 
$\tau$ to be $\tau^{(p)}$, the space of smooth sections 
$\Gamma^\infty E_\tau$ of $E_\tau$ is just the space of smooth 
$p-$ forms on $X$. We can therefore apply the version of the trace 
formula developed by Fried in Ref. 17. First we set up some 
more notation.
For $\sigma_j$ the natural representation of $SO(2k-1)$ on 
$\Lambda^j {\Bbb C}^{2k-1}$ one has the corresponding 
Harish--Chandra--Plancherel density given, for a suitable normalization 
of Haar measure $dx$ on $G$, by 

\begin{equation}\label{07}%%%%%%%
\mu_{\sigma_p(r)}=
\frac{\pi}{2^{4k-4}[\Gamma(k)]^2}
\left(
\begin{array}{c}
2k-1\\ p
\end{array}
\right)
rP_{\sigma_p}(r)\tanh(\pi r)\;,
\end{equation}
\\
for $0\le p \le k-1$, where 

\begin{equation}\label{08}%%%%%%%
P_{\sigma_p}(r)=\prod_{\ell=2}^{p+1}
\left[ 
r^2+\left(k-\ell+\frac{3}{2}\right)^2
\right]
\prod_{\ell=p+2}^{k}
\left[
r^2+\left(k-\ell+\frac{1}{2}\right)^2
\right]\;
\end{equation}
\\
is an even polynomial of degree $2k-2$. One has that $P_{\sigma_p}(r)=
P_{\sigma_{2k-1-p}}(r)$ and 
$\mu_{\sigma_p}(r)=\mu_{\sigma_{2k-1-p}}(r)$ 
for $k\le p\le 2k-1$. 
Define the Miatello coefficients \cite{Miatello} 
$a_{2\ell}^{(p)}$ for $G=SO_1(2k+1, 1)$ by 

\begin{equation}\label{09}%%%%%%%
P_{\sigma_p}(r)=\sum_{\ell=0}^{k-1}a_{2\ell}^{(p)}r^{2\ell}\;,
\qquad 0\le p\le 2k-1\;.
\end{equation}
\\
\\ 
Let ${\rm Vol}(\Gamma\backslash G)$ will denote the 
integral of the constant function $\mathbf{1}$ on $\Gamma\backslash G$ 
with respect to the $G-$ invariant measure on $\Gamma\backslash G$ 
induced by $dx$.
For $0\leq p\leq n-1$ the Fried trace formula applied to kernel
${\mathcal K}_t$ holds\cite{Fried}:

\begin{equation}
{\rm Tr}\left(e^{-t{\frak L}_{p}}\right)=I_{\Gamma}^{(p)}({\mathcal 
K}_t)
+I_{\Gamma}^{(p-1)}({\mathcal K}_t)
+H_{\Gamma}^{(p)}({\mathcal K}_t)+
H_{\Gamma}^{(p-1)}({\mathcal K}_t)
\mbox{,}
%\eqno{(3.1)}
\end{equation}
\\
where $I_{\Gamma}^{(p)}({\mathcal K}_t)\,, H_{\Gamma}^{(p)}({\mathcal 
K}_t)$ are identity and hyperbolic orbital integral respectively: 

\begin{equation}
I_{\Gamma}^{(p)}({\mathcal K}_t)
\stackrel{def}{=}\frac{\chi(1){\rm Vol}
(\Gamma\backslash G)}{4\pi}
\int_{\Bbb R}\mu_{\sigma_p}(r)e^{-t(r^2+p+\rho_0^2)}dr
\mbox{,}
\end{equation}

\begin{equation}
H_{\Gamma}^{(p)}({\mathcal K}_t)
\stackrel{def}{=}\frac{1}{\sqrt{4\pi t}}
\sum_{\gamma\in C_
\Gamma-\{1\}}\frac{\chi(\gamma)}{j(\gamma)}t_\gamma C(\gamma)
\chi_{\sigma_p}
(m_\gamma)
\exp\left\{-t(\rho_0^2+p)-
t_\gamma^2/(4t)\right\}
\mbox{,}
\end{equation}
\\
with $\rho_0=(n-1)/2$, and $\chi_\sigma(m)={\rm Tr}\;\sigma(m)$ 
for $m\in SO(2n-1)$. 
\\

For $p\geq 1$ there is a measure $\mu_{\sigma}(r)$ corresponding to a 
general irreducible representation $\sigma$ of ${\frak M}$. 
Let $\sigma_p$ be the 
standard representation of ${\frak M}=SO(n-1)$ on $\Lambda^p{\Bbb C}^{(n-1)}$. 
If
$n=2k$ is even then $\sigma_p\,\,(0\leq p\leq n-1)$ is always irreducible; if
$n=2k+1$ then every $\sigma_p$ is irreducible except for $p=(n-1)/2=k$, in 
which case $\sigma_k$ is the direct sum of two spin--$(1/2)$ representations 
$\sigma^{\pm}:\,\,\sigma_k=\sigma^{+}\oplus\sigma^{-}$. 
For $p=k$ the representation $\tau_k$ of $K=SO(2k)$ on 
$\Lambda^k {\Bbb C}^{2k}$ is not irreducible, 
$\tau_k=\tau_k^{+}\oplus\tau_k^{-}$ is the direct sum of spin--$(1/2)$ 
representations.

\subsection{Case of the trivial representation}

For $p=0$ (i.e. for smooth
functions or smooth vector bundle sections) the measure 
$\mu(r)\equiv \mu_{0}(r)$ corresponds to the trivial representation of 
${\frak M}$. 
Therefore we take $I_{\Gamma}^{(-1)}({\mathcal K}_t)
=H_{\Gamma}^{(-1)}({\mathcal K}_t)=0$. 
Let $\chi_{\sigma}(m) = {\rm trace}(\sigma(m))$ be the character of 
$\sigma$,
for $\sigma$ a finite--dimensional representation of ${\frak M}$.
Since $\sigma_0$ is the 
trivial representation one has
$\chi_{\sigma_0}(m_{\gamma})=1$. In this case formula (30)
reduces exactly to the trace formula for $p=0$   
\cite{Wallach,Elizalde,Bytsenko0,Williams,Bytsenko1,Bytsenko2}, 

\begin{equation}
I_{\Gamma}^{(0)}({\mathcal K}_t)
=\frac{\chi(1)\mbox{vol}(\Gamma\backslash G)}
{4 \pi}\int_{\Bbb R}\mu_{\sigma_0}(r)e^{-t(r^2+\rho_0^2)}dr+
H_{\Gamma}^{(0)}({\mathcal K}_t)
\end{equation}
\\
The function 
$H_{\Gamma}^{(0)}({\mathcal K}_t)$ has the form

\begin{equation}
H_{\Gamma}^{(0)}({\mathcal K}_t)
=\frac{1}{\sqrt{4\pi t}}
\sum_{\gamma\in C_
\Gamma-\{1\}}\chi(\gamma)t_\gamma j(\gamma)^{-1}
C(\gamma)\exp \{-t\rho_0^2-t_\gamma^2/(4t)\}
\mbox{.}
\end{equation}
\\
\\

\subsection{Odd dimensional manifolds with cusps}

Taking into account the fixed Iwasawa decomposition $G=KAN$,
consider a $\Gamma-$cuspidal minimal parabolic subgroup $P$ of $G$
with the Langlands decomposition $P=BAN$, $B$ being the
centralizer of $A$ in $K$.
Let us define the Dirac operator ${\frak D}$, assuming a spin
structure for $\Gamma \backslash {\rm Spin}(2k+1, 1)/{\rm
Spin}(2k+1))$. The spin bundle $E_{\tau_s}$ is the locally
homogeneous vector bundle defined by the spin representation
$\tau_s$ of the maximal compact group ${\rm Spin}(2k+1)$. One can
decompose the space of sections of $E_{\tau_s}$ into two
subspaces, which are given by the half spin representations
$\sigma_{\pm}$ of ${\rm Spin}(2k)\subset {\rm Spin}(2k+1)$.
Let us
consider a family of functions ${\mathcal K}_t$ over $G= {\rm
Spin}(2k+1, 1)$, which is given by taking the local trace for the
integral kernel $\exp(-t{\frak D}^2)$ (or ${\frak D}\exp(-t{\frak
D}^2)$). The Selberg trace formula has the form \cite{Park}:

$$
\!\!\!\!\!\!\!\!\!\!\!\!\!\!\!\!\!\!\!\!\!\!\!\!\!\!\!\!\!\!
\!\!\!\!\!\!\!\!\!\!\!\!\!\!\!\!\!\!\!\!\!\!\!\!\!\!\!\!\!\!
\!\!\!\!\!\!\!\!\!
\sum_{\sigma = \sigma_{\pm}}\sum_{\lambda_k\in \sigma_p}
{\hat {\mathcal K}}_t(\sigma, i\lambda_k)
-\frac{d(\sigma_{\pm})}{4\pi}\int_{\Bbb R}
{\rm Tr}\left(S_{\Gamma}(\sigma_{\pm},-i\lambda)
\right.
$$

\begin{equation}
\left.
\times
(d/ds)S_{\Gamma}(\sigma_{\pm},s)|_{s=i\lambda}\pi_{\Gamma}
(\sigma_{\pm},i\lambda)
({\mathcal K}_t)\right)d\lambda
= I_{\Gamma}({\mathcal K}_t) + H_{\Gamma}({\mathcal K}_t)
+ U_{\Gamma}({\mathcal K}_t)\,
\mbox{,}
\end{equation}
\\
where $\sigma_p \in {\rm Spec}\,{\frak D}$, $d(\sigma_{\pm})$ is the
degree of the half spin representation of ${\rm Spin}(2k)$,
$S_{\Gamma}(\sigma_{\pm}, i\lambda)$ is the scattering matrix and
$U_{\Gamma}({\mathcal K}_t)$ is an 
unipotent orbital integral. The analysis of the
unipotent orbital integral $U_{\Gamma}({\mathcal K}_t)$ gives the
following result \cite{Barbasch,Park}: all of the unipotent terms vanish
in the Selberg trace formula applied to the odd kernel
function. This means that using the Fried result \cite{Fried}
one obtains a result in the case of cusps
similar to that in the case of compact odd dimensional manifolds.

\section{The Spectral Functions on $p-$ Forms}

If $\{\lambda^{(p)}\}_{\ell=0}^{\infty}$ are eigenvalues of the 
operator 
$\delta d$ restricted on $p-$ forms and 
$\{\nu^{(p)}\}_{\ell=0}^{\infty}$ are
the eigenvalues of $d\delta$, then 
$\lambda_{\ell}^{(p)}=\nu_{\ell}^{(p+1)}$
with equal multiplicity. 
There are two equivalent eigenvalues problems

\begin{equation}
{\frak L}_p\omega_p = \lambda \omega_p \Longleftrightarrow
\begin{array}{l}
{\frak L}_{p+1}\,d\omega_p=\lambda\,d\omega_p\\
{\frak L}_{p-1}\,\delta\omega_p=\lambda\,\delta\omega_p
\end{array}
\mbox{.}
%\eqno{(12.33)}
\end{equation}
\\
This means that the spectra of the Hodge Laplacian acting on exact $p-$ 
forms
and on co--exact $(p-1)-$ forms are the same. The transverse part of 
the
skew symmetric tensor is represented by the co--exact $p-$ form
$\omega_p^{(CE)}=\delta\omega_{p+1}$, which trivially satisfies 
$\delta\omega_p^{(CE)}=0$,
and we denote by ${\frak L}_p^{(CE)} =\delta d$ the restriction of the 
Laplacian on the co--exact $p-$ form.

The goal now is to extract the co--exact $p-$ form on manifold 
which describes
the physical degrees of freedom of the system.
Choosing a basis $\{\omega_p^{\ell}\}$ of $p-$ forms (eigenfunctions of 
the Laplacian) we get for an arbitrary suitable function $F({\frak 
L}_p)$ 
\cite{Bytsenko}:

$$
\sum_{\ell} \:\langle\omega_p^\ell, F({\frak L}_p)\omega_p^\ell\rangle=
\sum_\ell \:\langle\omega_p^{\ell(CE)}, F({\frak 
L}_p)\omega_p^{\ell(CE)}\rangle
$$

\begin{equation}
+\sum_\ell \:\langle d\omega_{p-1}^\ell, F({\frak 
L}_p)d\omega_{p-1}^\ell\rangle
+b_pF(0)
\mbox{,}
\end{equation}
\\
where $b_p$ are the Betti numbers,  
$b_p\equiv b_p(M)= {\rm rank}_{\Bbb Z}H_p(M;{\Bbb Z})$. 
Using the properties of the Hodge Laplacian one can obtain

$$
\sum_\ell \:\langle d\omega_{p-1}^\ell, F({\frak 
L}_p)d\omega_{p-1}^\ell\rangle
=\sum_\ell \:\langle\delta\,d\omega_p^\ell, F({\frak L}_{p-1})
\delta\,d\omega_p^\ell\rangle
$$

\begin{equation}
=\sum_\ell \:\langle\omega_{p-1}^\ell, F({\frak 
L}_{p-1})\omega_{p-1}^\ell\rangle
-\sum_\ell \:\langle d\omega_{p-2}^\ell, F({\frak L}_{p-1})d
\omega_{p-2}^\ell\rangle -b_{p-1}F(0)
\mbox{.}
\end{equation}
\\
Finaly we get

\begin{equation}
{\rm Tr} F({\frak L}_p^{(CE)}) = \sum_{j=0}^{p} (-1)^j
({\rm Tr} F({\frak L}_{p-j})) - F(0)\sum_{j=0}^p(-1)^jb_{p-j}
\mbox{.}
%\eqno{(12.36)}
\end{equation}

Making the choice $F(x)=\exp (-tx)$, one can rewrite Eq. (39) using the
Fried's result (30) in the form

$$
{\rm Tr}\left(e^{-t{\frak L}_{p}^{(CE)}}\right)=
\sum_{j=0}^p(-1)^j\left(
I_{\Gamma}^{(p-j)}({\mathcal K}_t)
+I_{\Gamma}^{(p-j-1)}({\mathcal K}_t)
\right.
$$

\begin{equation}
\,\,\,\,\,\,\,\,\,\,\,\,\,\,\,\,\,\,\,\,\,\,\,\,\,\,\,\,\,\,\,\,
\,\,\,\,\,\,\,\,\,\,\,\,\,\,\,\,
\left.
+H_{\Gamma}^{(p-j)}({\mathcal K}_t)+
H_{\Gamma}^{(p-j-1)}({\mathcal K}_t)-b_{p-j}\right)
\mbox{.}
\end{equation}

%\section{Concluding remarks}

%These results could be useful, for example,
%in a special case of field theory for which the path integrals involve
%topological quantities (the Euler character, Witten index,
%etc.). In this case supersymmetry implies the existence of an equivariant 
%cohomology structure, which localizes the path integral by way of modes or
%characteristic classes \cite{Duistermaat,Berline,Szabo,Bytsenko1}.
%Usually
%equivariant cohomology is associated with supersymmetric theories, but
%it has very general applications. 

\section*{Acknowledgements}

 The work of A. A. Bytsenko was supported in part by the Russian
Foundation for Basic Research (grant No. 01-02-17157).

\end{document}